%% file: main.tex
\documentclass[conference]{IEEEtran}

\input{packages}
\input{definitions}

\begin{document}

%
%
%
\title{Inverter-Based Differential Amplifiers With Back-Gate Feedback Linearization \\
}

\author{\IEEEauthorblockN{Eric Danson and Jeffrey Walling}
\IEEEauthorblockA{\textit{Bradley Department of Electrical and Computer Engineering} \\
  Virginia Tech, Blacksburg, VA 24061, USA \\
  \{ecdanson, jswalling\}@vt.edu}
}

\maketitle%

%
%
\begin{abstract}

Feeding the common-source amplifier output to the back-gate terminal in \ac{FD-SOI} technology exploits the linearizing effect of negative feedback.
Analysis and simulation results in \qty{22}{\nano\meter} \ac{FD-SOI} show that back-gate feedback sets the overall gain approximately independent of the load, contributes no additional noise, and improves linearity by the back-gate voltage gain.
\Ac{IP3} enhancement is at least 60\(\times \) compared to without feedback in inverter-based, or complementary common-source, differential amplifiers.

\end{abstract}

\acresetall%


\acresetall%

%
%
%
\section{Introduction}

Nonlinear distortion reduction has applications in modern \ac{RF} systems that are limited by linearity rather than noise.
While massive \ac{MIMO} and phased arrays offer noise and interference suppression through spatial filtering, \ac{RF} spectrum crowding with interferers on adjacent channels causes distortion and limits the maximum signal power that \ac{RF} receivers can tolerate.

The back-gate terminal in \ac{FD-SOI} technology has been used for feedback to create closed-loop circuits.
Kuzmicz demonstrated back-gate feedback as a linearization technique for a two-stage operational transconductance amplifier operating in weak inversion~\cite{kuzmicz2021}.
Then, Weinreich and Murmann demonstrated that feedback linearization is not limited to weak inversion and works into strong inversion~\cite{weinreich2022}.
Weinreich and Murmann also measured a single-ended common-source amplifier to show that device variation is tightly controlled by the manufacturing process~\cite{weinreich2022}, so back-gate feedback is suitable for practical amplifiers requiring matched devices.

Differential amplifiers reject common-mode signals and suppress even-order harmonics.
Inverter-based, or complementary common-source, amplifiers provide high gain and higher power efficiency compared to a single-transistor common-source amplifier.
The same current that biases one transistor is reused for the complementary transistor, and both transistors contribute to the overall gain~\cite{sharroush2019}.
Cao \textit{et~al.} compared various inverter-based differential amplifiers using the \ac{NEF} figure of merit to show competitive power efficiency and noise performance relative to other topologies~\cite{cao2021}.
We propose combining inverter-based differential amplifiers with back-gate feedback linearization as a step towards overcoming linearity constraints in \ac{RF} systems, such as those used for telecommunications.

This paper is organized as follows.
\Cref{sec:background} provides background information.
\Cref{sec:ccs_analysis} provides a thorough analysis of the core circuit, followed in \cref{sec:ccs_diff_analysis} by analysis of differential configurations.
\Cref{sec:results} contains simulation results.
Finally, \cref{sec:conclusion} ends with conclusions.

\section{Background%
  \label{sec:background}}

\begin{figure}
  \hfill
  \begin{subfigure}{0.35\columnwidth}
    \adjustimage{max size={\textwidth}{2cm}, scale=1, center}{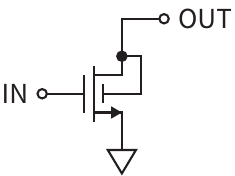}
    \caption{}
  \end{subfigure}
  \hfill
  \begin{subfigure}{0.55\columnwidth}
    \adjustimage{max size={\textwidth}{2cm}, scale=1, center}{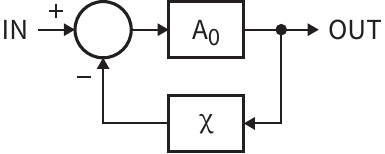}
    \caption{}
  \end{subfigure}
  \hfill
  \caption{Intrinsic gain stage (a) with back-gate feedback and (b) block diagram representation.%
    \label{fig:intrinsic_backgate_feedback}}
\end{figure}

Feeding the transistor output to the back gate in \ac{FD-SOI} technology exploits the linearizing effect of negative feedback (\cref{fig:intrinsic_backgate_feedback}) with a closed-loop voltage gain
\begin{equation}\label{eq:av_closed}
  A\_{v} = \frac{A_{0}}{1 + \chi A_{0}} = -\frac{g\_{m}r\_{o}}{1 - \chi g\_{m}r\_{o}}\rlap{\,.}
\end{equation}
The feedback factor is the back-gate capacitive coupling ratio
\begin{equation}\label{eq:coupling_ratio}
  \chi = -\frac{g\_{mb}}{g\_{m}} = -\frac{C\_{box}\parallel C\_{Si}}{C\_{ox}}
\end{equation}
where \(C\_{box} \), \(C\_{Si} \), and \(C\_{ox} \) are the buried oxide, thin silicon channel, and gate oxide capacitances, respectively~\cite{agarwal2020}.
When \(\left|\chi g\_{m}r\_{o} \right| \gg 1 \), the closed-loop gain is approximately \(1/\chi = -g\_{m}/g\_{mb} \) and is determined by process parameters.

Body feedback in bulk technologies does not provide the same linearization as in \ac{FD-SOI}.
The body coupling ratio is \(C\_{js}/C\_{ox} \) where \(C\_{js} \) is the junction capacitance between the substrate and source, and the depletion region width is inversely proportional to the doping concentration.
Steep retrograde body doping profiles (low doping concentration in the channel and high doping concentration in the body) reduce the depletion region width sensitivity to reverse body bias compared to a uniformly doped substrate~\cite{hu2009}.
With reverse body bias, the maximum depletion region width is constrained by the retrograde interface, but with forward body bias, the depletion region width decreases in the channel.
Whereas the buried oxide width and consequently \(C\_{box} \) are fixed in \ac{FD-SOI}, \(C\_{js} \) under forward bias in bulk technologies depends on the applied body voltage, making the body coupling ratio variable.

%
%
%
\section{Complementary Common-Source Analysis%
  \label{sec:ccs_analysis}}



\subsection{Gain}

\begin{figure}
  \hfill
  \begin{subfigure}{0.45\columnwidth}
    \adjustimage{max size={\textwidth}{3cm}, scale=1, center}{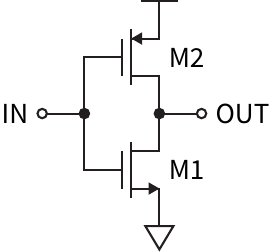}
    \caption{\label{fig:ccs_open}}
  \end{subfigure}
  \hfill
  \begin{subfigure}{0.45\columnwidth}
    \adjustimage{max size={\textwidth}{3cm}, scale=1, center}{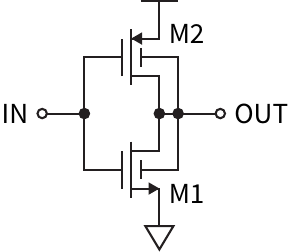}
    \caption{\label{fig:ccs_backgate}}
  \end{subfigure}
  \hfill
  \caption{Complementary common-source stage (a) in open loop and (b) with back-gate feedback.%
    \label{fig:ccs_open_backgate}}
\end{figure}

\begin{figure}
  \adjustimage{max size={\figwidth}{\textheight}, scale=1, center}{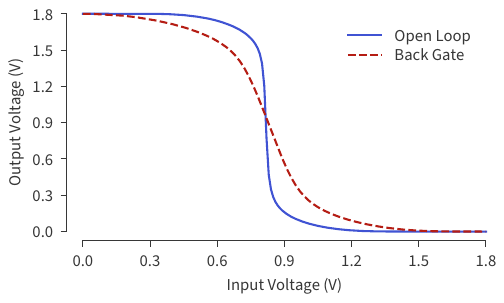}
  \caption{Inverter large-signal behavior in open loop and with back-gate feedback in FD-SOI technology.%
    \label{fig:large_signal_sweep_vgs}}
\end{figure}

Back-gate feedback makes the gain approximately independent of the load resistance.
Feedback changes the output resistance but not the input conductance.
The small-signal voltage gain of the complementary common-source amplifier in open loop
\begin{equation}
  A\_{v,ol} = -\left(g\_{mn} + g\_{mp} \right)\left(r\_{on}\parallel r\_{op} \right)
\end{equation}
and with back-gate feedback
\begin{equation}
  A\_{v,bg} = -\frac{\left(g\_{mn} + g\_{mp} \right)\left(r\_{on}\parallel r\_{op} \right)}{1 + \left(g\_{mbn} + g\_{mbp} \right)\left(r\_{on}\parallel r\_{op} \right)} \approx -\frac{g\_{mn} + g\_{mp}}{g\_{mbn} + g\_{mbp}}
\end{equation}
shows the same intrinsic gain through the front gate (\cref{fig:ccs_open_backgate}).
Where each configuration differs is the feedback reducing the output resistance by the back-gate voltage gain, so the overall gain is approximately the front- and back-gate conductance ratio \(1/\chi \) from \cref{eq:coupling_ratio} in saturation (\cref{fig:large_signal_sweep_vgs}).

\subsection{Noise}

The input-referred noise power spectral density of the complementary common-source amplifier in open loop and with back-gate feedback
\begin{equation}
  \begin{aligned}
    \frac{\overline{v\_{in}^{2}}}{\Delta f} &= 4kT\frac{\gamma\_{n}g\_{mn} + \gamma\_{p}g\_{mp}}{{\left(g\_{mn} + g\_{mp} \right)}^{2}} \\
    &+ \frac{1}{{\left(g\_{mn} + g\_{mp} \right)}^{2}C\_{ox}f}\left[\frac{g\_{mn}^{2}K\_{n}}{{\left(WL \right)}\_{n}} + \frac{g\_{mp}^{2}K\_{p}}{{\left(WL \right)}\_{p}} \right]
  \end{aligned}
\end{equation}
suggests that feedback contributes no additional noise (\cref{fig:ccs_open_backgate}).
Compared to a single common-source amplifier with a current source load, complementary amplifiers have lower input-referred noise because both transistors contribute to the signal gain.
Noise power spectral density is doubled for differential configurations because there are twice as many transistors to provide the same gain.
Without a current source to set the operating point, feedback forward biases the transistor bodies and lowers the threshold voltage.
The increased drain current compared to open loop with the same transistor sizes would shift the transistors' operating points and associated noise parameters.
Therefore, feedback should not contribute noise if the transistors' operating points are kept constant.

\subsection{Linearity}

Linearity enhancement by nonlinear distortion reduction is back-gate feedback's primary advantage.
Expanding the drain-to-source current of each complementary common-source transistor in open loop (\cref{fig:ccs_open}) as a Taylor series approximation around the operating point \(\left(V\_{GS},\, V\_{DS},\, I\_{DS} \right) \),
\begin{equation}
  \begin{aligned}
    i\_{ds}\left(v\_{gs}, v\_{ds} \right) &= g\_{m1}v\_{gs} + g\_{ds1}v\_{ds} \\
    &+ g\_{m2}v\_{gs}^{2} + x\_{11}v\_{gs}v\_{ds} + g\_{ds2}v\_{ds}^{2} \\
    &+ g\_{m3}v\_{gs}^{3} + x\_{12}v\_{gs}v\_{ds}^{2} + x\_{21}v\_{gs}^{2}v\_{ds} + g\_{ds3}v\_{ds}^{3} + \cdots
  \end{aligned}
\end{equation}
where
\begin{equation}
  \begin{gathered}
    g_{\mathrm{m}k} = \frac{1}{k!}\frac{\partial^{k} i\_{DS}}{\partial v\_{GS}^{k}}\,,\quad g_{\mathrm{ds}k} = \frac{1}{k!}\frac{\partial^{k} i\_{DS}}{\partial v\_{DS}^{k}}\,, \\
    x_{pq} = \frac{1}{p!q!}\frac{\partial^{p + q} i\_{DS}}{\partial v\_{GS}^{p} \partial v\_{DS}^{q}}
  \end{gathered}
\end{equation}
are the derivatives evaluated at the operating point~\cite{blaakmeer2008}.
The drain-to-source currents through each transistor
\begin{equation}\label{eq:idsnp}
  i\_{dsn}\left(v\_{gs}, v\_{ds} \right) = -i\_{dsp}\left(v\_{gs}, v\_{ds} \right)
\end{equation}
are equal and opposite in magnitude.
Relating the output voltage \(\left(v\_{ds} \right) \) to the input voltage \(\left(v\_{gs} \right) \) by a power series
\begin{equation}
  v\_{ds} = a_{1}v\_{gs} + a_{2}v\_{gs}^{2} + a_{3}v\_{gs}^{3} + \cdots\rlap{\,,}
\end{equation}
substituting into \cref{eq:idsnp}, and equating coefficients without cross derivatives for simplicity, we find
\begin{equation}
  \begin{aligned}
    a_{1} &= -G\_{m1}G\_{ds1}^{-1} \\
    a_{2} &= -\left(G\_{m2} + G\_{ds2}a\_{1}^{2} \right)G\_{ds1}^{-1} \\
    a_{3} &= -\left(2G\_{ds2}a_{1}a_{2} + G\_{m3} + G\_{ds3}a_{1}^{3} \right)G\_{ds1}^{-1}
  \end{aligned}
\end{equation}
where
\begin{equation}
  \begin{aligned}
    G_{\mathrm{m}k} &= g_{\mathrm{mn}k} + g_{\mathrm{mp}k} \\
    G_{\mathrm{ds}k} &= g_{\mathrm{dsn}k} + g_{\mathrm{dsp}k} \\
  \end{aligned}
\end{equation}
are the combined conductances of each transistor.
Analysis with back-gate feedback (\cref{fig:ccs_backgate}) likewise yields coefficients
\begin{equation}
  \begin{aligned}
    a_{1} &= -G\_{m1}{\left(G\_{mb1} + G\_{ds1} \right)}^{-1} \\
    a_{2} &= -\left[G\_{m2} + \left(G\_{mb2} + G\_{ds2} \right)a\_{1}^{2} \right]{\left(G\_{mb1} + G\_{ds1} \right)}^{-1} \\
    a_{3} &= -\left[2\left(G\_{mb2} + G\_{ds2} \right)a_{1}a_{2} + G\_{m3} \right. \\
    &\qquad + \left.\left(G\_{mb3} + G\_{ds3} \right)a_{1}^{3} \right]{\left(G\_{mb1} + G\_{ds1} \right)}^{-1}
  \end{aligned}
\end{equation}
by replacing \(g_{\mathrm{ds}k} \) with \(\left(g_{\mathrm{mb}k} + g_{\mathrm{ds}k} \right) \) for the output conductance.
The relative linearity enhancement for back-gate feedback compared to open loop, considering \ac{IP3} defined as \(\mathrm{IP3} = {\left[\left(4/3 \right)a_{1}/a_{3} \right]}^{1/2} \),
\begin{equation}
  \begin{aligned}
    \frac{\mathrm{IP3}\_{bg}}{\mathrm{IP3}\_{ol}} &\propto {\left(\frac{G\_{mb1} + G\_{ds1}}{G\_{ds1}} \right)}^{2} \\
    &= {\left[1 + \left(g\_{mbn1} + g\_{mbp1} \right)\left(r\_{on1}\parallel r\_{op1} \right) \right]}^{2}
  \end{aligned}
\end{equation}
suggests that transistors with longer channel lengths and larger output resistances provide better linearity.

\section{Differential Amplifier Analysis%
  \label{sec:ccs_diff_analysis}}

\begin{figure}
  \centering
  \begin{subfigure}{0.7\columnwidth}
    \adjustimage{max size={\textwidth}{\textheight}, scale=1, center}{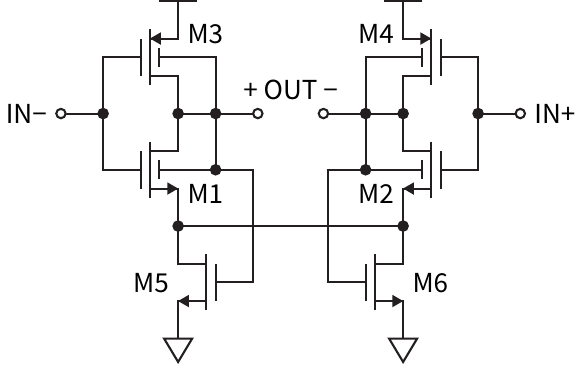}
    \caption{\label{fig:ccs_backgate_diff_scmfb}}
  \end{subfigure}
  \par\vspace{0.5\baselineskip}
  \begin{subfigure}{0.7\columnwidth}
    \adjustimage{max size={\textwidth}{\textheight}, scale=1, center}{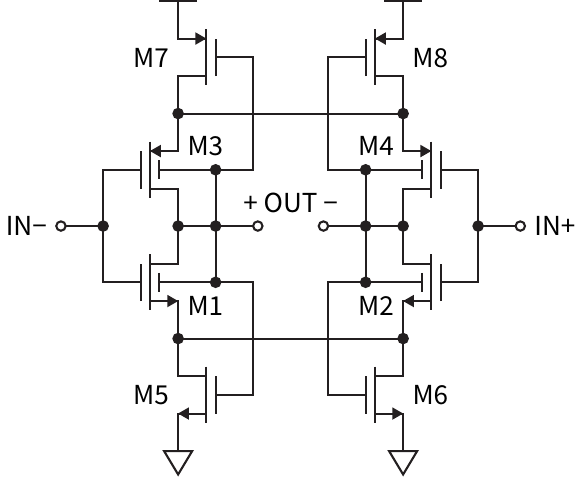}
    \caption{\label{fig:ccs_backgate_diff_dcmfb}}
  \end{subfigure}
  \caption{(a) Single and (b) dual common-mode feedback (SCMFB and DCMFB, respectively) differential amplifiers.%
    \label{fig:ccs_backgate_diff}}
\end{figure}

\subsection{Biasing}

We consider two differential amplifiers: one with \ac{SCMFB} and another with \ac{DCMFB} (\cref{fig:ccs_backgate_diff}).
\Ac{CMFB} transistors \(M_{5} \) to \(M_{8} \) act as current sources, so the differential pair and \ac{CMFB} transistor size ratios set the transconductance efficiency \(g\_{m}/I\_{D} \).
With dual \ac{CMFB}, the drain current
\begin{equation}
  I\_{D} \propto {\left[\mu\_{n}{\left(\frac{W}{L} \right)}_{5,6} \right]}^{1/2}\parallel {\left[\mu\_{p}{\left(\frac{W}{L} \right)}_{7,8} \right]}^{1/2}\rlap{\,.}
\end{equation}
Increasing the \ac{CMFB} transistor sizes increases \(I\_{D} \) and reduces the input transistors' \(g\_{m}/I\_{D} \).
Conversely, increasing the input transistor sizes keeps \(I\_{D} \) constant and increases \(g\_{m}/I\_{D} \).

\subsection{Common-Mode Rejection}

The small-signal common-mode voltage gain with single \ac{CMFB}
\begin{equation}
  A\_{v,cm} \approx -\frac{g\_{m3,4}}{g\_{m5,6}}
\end{equation}
and with dual \ac{CMFB}
\begin{equation}
  A\_{v,cm} \approx -\frac{1}{\left(g\_{m5,6} + g\_{m7,8} \right)\left(r\_{o5,6}\parallel r\_{o7,8} \right)}
\end{equation}
are approximately the same in open loop and with back-gate feedback.
The differential-mode gain for each amplifier is the same as that of the complementary common source.
Therefore, the dual \ac{CMFB} amplifier has a greater \ac{CMRR} by a factor of
\begin{equation}
  \frac{\mathrm{CMRR}\_{DCMFB}}{\mathrm{CMRR}\_{SCMFB}} \approx \frac{g\_{m3,4}}{g\_{m5,6}}\left(g\_{m5,6} + g\_{m7,8} \right)\left(r\_{o5,6}\parallel r\_{o7,8} \right)
\end{equation}
but has larger output capacitance and requires greater voltage headroom to keep all the transistors operating in saturation.

%
%
\section{Simulation Results%
  \label{sec:results}}

All the simulation results are in GlobalFoundries' \qty{22}{\nano\meter} \ac{FD-SOI} technology using \qty{1.8}{\volt} thick-oxide I/O devices to extend the linear operating range over that of the \qty{0.8}{\volt} thin-oxide devices.

\begin{figure}
  \adjustimage{max size={\figwidth}{\textheight}, scale=1, center}{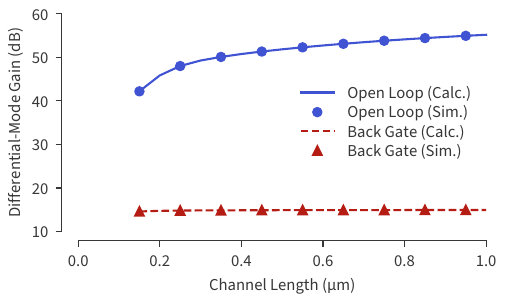}
  \caption{Calculated and simulated differential-mode gain.%
    \label{fig:dcmfb_dmgain_sweep_len}}
\end{figure}

\begin{figure}
  \adjustimage{max size={\figwidth}{\textheight}, scale=1, center}{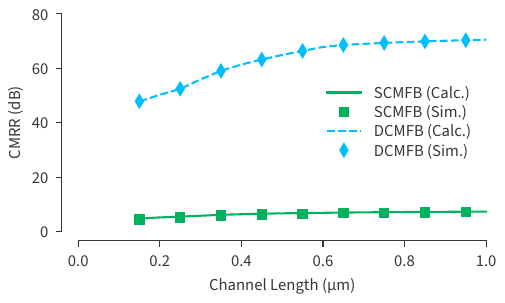}
  \caption{Calculated and simulated common-mode rejection ratio with back-gate feedback.%
    \label{fig:cmrr_sweep_len}}
\end{figure}

Back-gate feedback sets the differential-mode gain to about five, according to the back-gate coupling ratio, and is approximately independent of channel length compared to open loop (\cref{fig:dcmfb_dmgain_sweep_len}).
Feedback voltage division or additional amplifier stages may be used for higher gain depending on bandwidth requirements.
Dual \ac{CMFB} greatly increases the \ac{CMRR} by \qty{43}{\decibel} at \(L = \qty{150}{\nano\meter} \) and by \qty{63}{\decibel} at \(L = \qty{1}{\micro\meter} \) compared to the single \ac{CMFB} configuration (\cref{fig:cmrr_sweep_len}).
Monte Carlo simulation with 100 points each shows up to \qty{0.4}{\decibel} standard deviation in \ac{CMRR} due to \ac{CMFB} transistor mismatch (\cref{tab:cmrr_stats}), so there is no significant performance degradation across corners.

\begin{table}
  \centering
  \caption{Back-Gate Feedback CMRR%
    \label{tab:cmrr_stats}}
  \begin{booktabs}{
    colspec = {cccc},
    cell{1}{2} = {r=2}{m}, 
    cell{3}{1} = {r=2}{m},
    cell{5}{1} = {r=2}{m},
    cell{1}{3} = {c=2}{m} 
  }
    \toprule
    & \textbf{Length (\unit{\micro\meter})} & \textbf{CMRR (\unit{\decibel})} & \\ \cmidrule[lr]{3-4}
    & & Mean & Std. Dev. \\ \midrule
    SCMFB & 0.15 & 4.43 & 0.0888 \\
    & 1 & 7.15 & 0.115 \\ \addlinespace[0.5em]
    DCMFB & 0.15 & 47.9 & 0.344 \\
    & 1 & 70.5 & 0.404 \\ \bottomrule
  \end{booktabs}
\end{table}

\begin{figure}
  \adjustimage{max size={\figwidth}{\textheight}, scale=1, center}{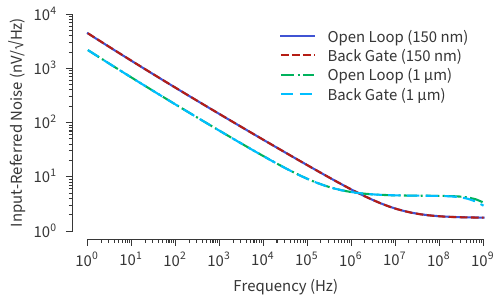}
  \caption{Simulated differential amplifier input-referred noise density at \qty{150}{\nano\meter} and \qty{1}{\micro\meter} channel lengths with \(g\_{m}/I\_{D} = \qty{15}{\siemens\per\ampere} \) (moderate inversion).%
    \label{fig:dcmfb_noise_sweep_freq}}
\end{figure}

\begin{figure}
  \adjustimage{max size={\figwidth}{\textheight}, scale=1, center}{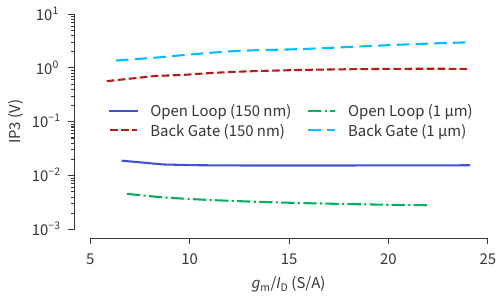}
  \caption{Simulated differential amplifier third-order intercept point at \qty{150}{\nano\meter} and \qty{1}{\micro\meter} channel lengths.%
    \label{fig:dcmfb_vip3_sweep_gmid}}
\end{figure}

Noise and linearity with back-gate feedback compares favorably.
The \ac{CMFB} transistors act as current sources to set the differential amplifier operating current.
At the same bias conditions, back-gate feedback consumes no extra power and contributes no additional noise compared to open loop (\cref{fig:dcmfb_noise_sweep_freq}).
The relative \ac{IP3} enhancement for back-gate feedback over open loop increases with channel length by 60\(\times \) at \(L = \qty{150}{\nano\meter} \) and by 700\(\times \) at \(L = \qty{1}{\micro\meter} \) because of larger output resistance (\cref{fig:dcmfb_vip3_sweep_gmid}).
Further linearity improvement is possible by using a cascode amplifier topology to increase the output resistance.
Though longer channel lengths decrease \ac{IP3} in open loop, the opposite is true with back-gate feedback.
Longer channel lengths increase \ac{IP3} as the open-loop gain \(A_{0} \) increases and makes the closed-loop gain closer to the inverse feedback factor \(1/\chi \) according to \cref{eq:av_closed}.

%
%
\section{Conclusion%
  \label{sec:conclusion}}

Back-gate feedback in \ac{FD-SOI} technology can realize gain that is approximately independent of load resistance while contributing no additional noise and improving linearity by a factor proportional to output resistance, based on analysis and simulation results.
Inverter-based, or complementary common-source amplifiers, as a core building block are advantageous for noise and power efficiency compared to single-transistor common-source amplifiers.
Inverter-based differential amplifiers, together with back-gate feedback, are a potential step towards overcoming linearity constraints in \ac{RF} systems.

%
%
\section*{Acknowledgment}

We would like to acknowledge GlobalFoundries' University Partnership Program for providing process development kits.

%
%
\bibliography{bibliography/IEEEabrv, bibliography/IEEEbstctl, bibliography/references}

\end{document}

%% file: packages.tex
\usepackage{acro}
\usepackage{adjustbox}
\usepackage{balance}
\usepackage{cite}

\AtBeginDocument{\bstctlcite{IEEEexample:BSTcontrol}}

\usepackage{etoolbox}
\usepackage[T1]{fontenc}
\usepackage{mathtools}
\usepackage{microtype}
\usepackage{mleftright}

\mleftright%

\usepackage{newtx}

\DeclareEncodingSubset{TS1}{ntxtlf}{0}

\usepackage{siunitx}

\usepackage{soul}

\makeatletter
\let\MYcaption\@makecaption%
\makeatother

\usepackage[font=footnotesize, labelformat=simple]{subcaption}

\makeatletter
\let\@makecaption\MYcaption%
\makeatother

\usepackage{tabularray}

\UseTblrLibrary{booktabs}

\AtEndPreamble{
  \usepackage[capitalize]{cleveref}

  \crefname{equation}{}{}
  \Crefname{equation}{Eq.}{Eqs.}
  \Crefname{figure}{Fig.}{Figs.}
}

%% file: definitions.tex
\DeclareAcronym{CMFB}{short=CMFB, long=common-mode feedback}
\DeclareAcronym{CMRR}{short=CMRR, long=common-mode rejection ratio}
\DeclareAcronym{DCMFB}{short=DCMFB, long=dual \acl{CMFB}}
\DeclareAcronym{FD-SOI}{short=FD-SOI, short-indefinite=an, long=fully depleted silicon on insulator}
\DeclareAcronym{IP3}{short=IP3, short-indefinite=an, long=third-order intercept point}
\DeclareAcronym{MIMO}{short=MIMO, long=multiple-input multiple-output}
\DeclareAcronym{NEF}{short=NEF, short-indefinite=an, long=noise efficiency factor}
\DeclareAcronym{RF}{short=RF, short-indefinite=an, long=radio frequency}
\DeclareAcronym{SCMFB}{short=SCMFB, short-indefinite=an, long=single \acl{CMFB}}

\newcommand*{\submathrm}[1]{\sb{\mathrm{#1}}}
\DeclareRobustCommand{\_}{\ifmmode\expandafter\submathrm\else\textunderscore\fi}

\newcommand*{\figwidth}{0.9\columnwidth}
